\documentclass[aps,preprint]{revtex4}
\usepackage{bm}
\usepackage{amsfonts}
\usepackage{amsmath}
\usepackage{amssymb}
\usepackage{graphicx}%
    \makeatletter
    
    \newcommand{\Rmnum}[1]{\expandafter\@slowromancap\romannumeral #1@}
    \makeatother
\allowdisplaybreaks[4]

\begin{document}
\preprint{CTP-SCU/2012003  }
\title{ Microscopic quantum structure of black hole and vacuum versus  quantum statistical origin of gravity}

\author{Shun-Jin Wang\\
        \normalsize\textit{Center for Theoretical Physics, College of Physical Science and Technology,}\\
        \normalsize\textit{Sichuan University, Chengdu 610064, China}\\
        \normalsize\textit{E-mail:}
        \texttt{sjwang@home.swjtu.edu.cn}
       }
\begin{abstract}
The Planckon densely piled model of vacuum is proposed. Based on this model, the microscopic quantum structure of Schwarzschild black hole and quantum statistical origin of its gravity are studied. The cutoff of black hole horizon leads to Casimir effect inside the horizon. This effect makes the inside vacuum has less zero quantum fluctuation energy than that of outside vacuum and the spin 1/2 radiation hole excitations are resulted inside the horizon. The mean energy of the radiation hole excitations is related to the temperature decrease of the Hawking-Unruh type by the period law of the Fermion temperature greens function and a temperature difference as well as gravity are created on the horizon. A dual relation of the gravity potentials between inside and outside regions of the black hole is found. An attractor behaviour of the horizon surface is unveiled. The gravity potential inside the black hole is linear in radial coordinate and no singularity exists at the origin of the black hole, in contrast to the conventional conjecture. All the particles absorbed by the black hole have fallen down to the horizon and converted into spin 1/2 radiation quanta with the mean energy related to the Hawking-Unruh temperature, the thermodynamic equilibrium and the mechanical balance make the radiation quanta be tightly bound in the horizon. The gravitation mass $2M$ and physical mass $M$ of the black hole are calculated. The calculated entropy of the black hole is well consistent with Hawking. Outside the horizon, there exist thermodynamic non-equilibrium and mechanical non-balance which lead to an outward centrifugal energy flow and an inward gravitation energy flow. The lost vacuum energy in the negative gravitation potential region has been removed to the black hole surface to form a spherical Planckon shell with the thickness of Planckon diameter so that energy conservation is guaranteed.
\end{abstract}
\maketitle
\tableofcontents

\section{Structure of vacuum: Plankon densely piled vacuum model}

Black holes are typical objects in astrophysics, which show full nature of gravity\cite{1,2,3,4,5}. In this article, we shall try to explore the microscopic quantum structure of vacuum and the quantum statistical origin of gravity by virtue of investigation of back holes.

During the search for microscopic foundation of the temperature as well as the four thermodynamic laws of black holes, it is naturally to conjecture that vacuum has a microscopic quantum structure and the thermodynamic laws of black holes are rooted in microscopic quantum statistical physics.

To make the above conjecture concrete, we have proposed a novel model of vacuum which is piled up densely by extremely tiny Planck radiation quantum spheres and is a kind of a liquid crystal under mean field and semi-classical approximation. The Planck radiation quantum sphere called Planckon, is made of localized standing radiation waves on average with radius $r_p\approx  10^{-33}cm$, mass $m_p\approx 10^{-5}g$, energy  $e_p\approx 10^{19}GeV$ , and spin $s_p={\hbar}/{2}$.

There are three ways to find the parameters of the Planckon. The first way requires that the Planckon is a black hole with its radius
$ r={2Gm}/{c^2}$ which is made of a quantum standing wave on the sphere with wave length $\lambda=4\pi r$, wave vector $k={2\pi}/{\lambda}={1}/{2r}$, quantized energy $e(r)=\hbar kc={\hbar c}/{2r}=\hbar\omega$, and mass $m(r)=e(r)/c^2=\hbar
/2rc$. From above requirements , one obtains radius, energy, mass, and spin of the Planckon as follows:
\begin{eqnarray}
r_p=\sqrt{\frac{G\hbar}{c^3}},\ e_p=\frac{\hbar c}{2r_p},\ m_p=\frac{1}{2}\sqrt{\frac{c\hbar}{G}},\ s_p=m_pcr_p=\frac{1}{2}\hbar
\end{eqnarray}
The second way requires that the Planckon is a quantum radiation sphere with the quantum spin $\hbar/2$  and the balance between gravitation force and centrifugal force is established. The third one requires that the Planckon is a quantum radiation standing wave on a sphere with wave length $\lambda=4\pi r$, and the gravitation force and centrifugal force are balanced. All the three approaches reach the same results. The Planckon is the smallest microscopic quantum black hole with spin\ $\hbar/2$ and made of localized radiation standing wave on the sphere with wave length $\lambda=4\pi r$. It is also the smallest microscopic quantum particle with the heaviest mass.

Summary of Planckon's parameters: scales of space-time $r_p=(\hbar G/c^3)^{1/2}$  and  \ $t_p=(\hbar G/c^5)^{1/2}$; volume $v_p={4\pi}r^3_p/{3}$; spin\ $s_p=\hbar/2$; mass\ $m_p=\frac{1}{2}({\hbar c}/{G})^{1/2}={\hbar}/{2cr_p}$, energy $e_p=m_pc^2={\hbar c}/{2r_p}=\hbar \omega_p$, wave nature $\omega_{p}={c}/{2r_p} ,\ k_p={e_p}/{\hbar c}={1}/{2r_p},\ \lambda_p=4\pi r_p$;
zero enegy:\ \ $e_{p0}=e_p/2={\hbar c}/{4r_p}$ ; zero energy density of Planckon: \\
$\rho_{p0}=e_{p0}/v_p=\frac{3}{16\pi}\frac{c^7}{G^2\hbar}=\frac{\rho_p}{2}\ (e_{p0}= \rho_{p0}v_{p}=\frac{\hbar c}{4r_{p}})$; vacuum zero energy
density is twice of that of Planckon (since vacuum is made of densely piled Planckons and each Planckon has two spin states):$\rho_v=\rho_p=2\rho_{p0}
=\frac{3}{8\pi}\frac{c^7}{G^2\hbar},\ \rho_v v_p=e_p=\frac{\hbar c}{2r_p}$ ; others: $Gm_p=r_pc^2/2$,\ $2Gm_p/c^2=r_p$,\ $G=6.6\times10^{-8}cgs$.

\textbf{Microcopic structure of Planckon vacuum:} there are two ways (phases)
to densely pile vacuum by spherical Planckons\cite{6}:\\
Face-center cubic crystal: pile layer order is  ABCABCABC$\cdots$ pile period is ABC\\
Hexagon  dense crystal:\ pile layer order is  ABABABABAB$\cdots$pile period is AB

For infinite crystal, the two phases may be degenerate in energy, since a rotation of  the C layer of the face-center cubic by $60^0$\ reaches the Hexagon dense phase.
However, for finite crystal, due to dislocation or defects,their energies become non-degenerate. For a cubic with side length $R$(for a sphere with radius R, the result is the same), the volume is $V=R^3$ , the smallest dislocation volume is
$\Delta V=3R^2\Delta R=6R^2r_p$  with the smallest $\Delta R=2r_p$ . Hence
\ $\frac{\Delta V}{V}\sim\frac{6r_p}{R}$ . For the mass density\ $\rho=\frac{m}{V}$, the ratio of the mass density change $\Delta\rho$ due to the dislocation $\Delta V $  to the mass density $ \rho $ is $\frac{\Delta\rho}{\rho}=\frac{6r_p}{R}$; for the vacuum energy density $\rho=\rho_p$, the dislocation energy and mass per Planckon on average is:\
$\Delta\rho v_p=Mc^2$, $e_p=\rho_pv_p=m_pc^2$, $\frac{M}{m_p}\sim\frac{6r_p}{R}$
; for the proton ($R_p\approx10^{-13}cm$ ), the dislocation mass is\ $M_p\sim\frac{6r_p}{R_p}m_p\approx10^{-24}g$.

Vacuum elasticity coefficient:\ $K\approx\frac{e_p}{r^{2}_{p}}\approx\frac{10^{19}GeV}{10^{-66}cm^2}\sim10^{82}erg/cm^2$

Vacuum transverse wave velocity $c_\nu$ equal to $c$:\ $c_{\nu}^{2}=\frac{\mu_\nu}{\rho_\nu/c^2}=\frac{\mu_\nu c^2}{\rho_\nu}=c^2$  (where\ $\rho_{\nu}/c^2$ is vacuum mass density) indicating that vacuum transverse stress\ $\mu_\nu$ is equal to vacuum energy density $\rho_\nu:\mu_\nu=\rho_\nu$.

Vacuum longitudinal stress $K_\nu$ : from $\rho_\nu\sim C/r_p^4 \propto C/v_p^{4/3}$, one gets $K_\nu=-v_ p\frac{\partial(\rho_\nu c^2)}{\partial v_p}=\frac{4}{3}\rho_\nu=\frac{4}{3}\mu_\nu$¡£

The solid property of vacuum comes from its extremely large mass density and stress, and its liquid property is due to the quantum fluctuation of the Planckon and the related variability of the localized radiation waves. Thus the Planckon densely piled vacuum is a kind of liquid crystal.

\section{Microscopic quantum structure of Schwarzschild black hole and quantum statistical origin of its gravity and temperature}

\subsection{ Quantum statistical meaning of the temperature on black hole horizon }
Basic parameters:

Balck hole radius
\begin{eqnarray}
r_h=\frac{2GM}{c^2}, \ g_{00}(r)=(1-\frac{r_h}{r})
\end{eqnarray}

Hawking-Unruh temperature\cite{4,5}
\begin{eqnarray}
T_h=\frac{\hbar \kappa}{2\pi k_B c}
\end{eqnarray} The gravity acceleration on the horizon
\begin{eqnarray}
\kappa=\lim_{r\rightarrow r_{h}}\sqrt{\frac{g^{rr}}{g_{00}}}\frac{dg_{00}(r)}{dr}=\frac{c^2}{r_h}
\end{eqnarray}
To explore the quantum statistical structure of the black hole and the origin of its gravity, we introduce the spherical standing radiation wave as vacuum excitation quantum as follows: wave length $\lambda_h=4\pi r_h$  and wave vector $k_h=\frac{2\pi}{\lambda_h}=\frac{1}{2r_h}$ , wave frequency $\omega_h=\frac{c}{2r_h}=2\pi\frac{c}{\lambda_h}=2\pi\nu_h$, quantum energy $e_h=\frac{\hbar c}{2r_h}=\hbar\omega_h$, quantum mass $m_h=e_h/c^2=\hbar/2cr_h$, and spin $s_h=m_hcr_h=\frac{\hbar}{2}$. The radiation quantum moving in spherical horizon is a spin 1/2 fermion which is an astronomic dual of Planckon as follows: $r_p\leftrightarrow r_h$\\
Planckon quantum:
\begin{eqnarray}
e_p=\frac{\hbar c}{2r_p},\ m_p=\frac{\hbar}{2cr_p},\ s_p=\frac{\hbar }{2}, \lambda_p=4\pi r_p
\end{eqnarray}
Radiation quantum:
\begin{eqnarray}
e_h=\frac{\hbar c}{2r_h},\ m_h=\frac{\hbar}{2cr_h},\ s_h=\frac{\hbar}{2},\ \lambda_h=4\pi r_h
\end{eqnarray}

According to the temperature Green's function for fermion\cite{7,8}, the horizon temperature $T_h$ and the corresponding thermal energy $k_B T_h$ are related to the radiation quantum mean energy $e_h$
\begin{eqnarray}
e_h/k_BT_h=\beta e_h=\pi\ (\beta=1/k_BT_h)
\end{eqnarray}
It turns to be
\begin{eqnarray}
\pi k_BT_h=e_h=\frac{\hbar c}{2r_h}=\frac{\hbar \kappa}{2c}\nonumber
\end{eqnarray}
or
\begin{eqnarray}
T_h=
\frac{\hbar \kappa}{2\pi ck_B},\ \kappa=\frac{2GM}{r_h^2}=\frac{c^2}{r_h}
\end{eqnarray}
which is exactly the Hawking-Unruh Formulae of the black hole temperature.

This is the first step to explore the microscopic contents of Hawking-Unruh formulae and find the elementary quantum constituent of the quantum statistical system of gravity--the radiation quantum and its motion mode, energy, wave vector, and spin, as well as the relation between the statistical temperature $T_h$ and the constituent mean energy $e_h$. The information is enough for a quantum statistical (ideal gas) system if the number ( or number density ) of the constituents is provided. In the following, we shall explore that the black hole temperature and the radiation quantum mean energy can be found from the Casimir effect-the cutoff effect of the black hole horizon.

It should be noted that in the black hole horizon and for radiation particles, there exists a mechanical balance
such that the gravity acceleration is equal to the centrifugal acceleration:
\begin{eqnarray}
\kappa=a_{gr}=\frac{2GM}{r_h^2}=\frac{c^2}{r_h}=a_{centrif}
\end{eqnarray}
This observation leads to another way to find the black hole radius. If it is required that the particles moving in the black hole horizon must be the radiation particles with the speed of light and their gravity acceleration must balance their centrifugal acceleration so that the particles are confined in the horizon, then $a_{gr}=\kappa=\frac{2GM}{r_h^2}=a_{centrif}=\frac{c^2}{r_h}$ which leads to the expression of the black hole radius\ $r_h=2GM/c^2$ . Since every particle in horizon is moving with the speed of light, the space-time metrices become singular in horizon,\ $g_{00}(r_h)=0$ and\ $g_{rr}(r_h)=\infty$. For the spherical neutral black hole, the possible solutions are\ $g_{00}(r_h)=1-\frac{r_h}{r}=1/g_{rr}(r)
=1-\frac{2GM}{rc^{2}}$.

\subsection{Microscopic structure of black hole and Casimir effect of spherical horizon cutoff}
In this subsection, we shall show that the black hole temperature and the radiation quantum mean energy can be found from the Casimir effect-the cutoff effect of the horizon.

Since particles can not escape from horizon, the spherical surface of the black hole horizon becomes a cutoff surface for radial waves inside of the sphere. The spin 1/2 radial radiation excitation wave function $u(r)$ of the Planckon vacuum medium according to the two-component Dirac equation has the following form and boundary condition at the spherical horizon surface of the black hole
\begin{eqnarray}
u(r)\xrightarrow{r\rightarrow r_h} sin(2kr)/2kr \nonumber\\
sin(2kr_h)=0,\ 2kr_h=n\pi,\ k_n =\frac{n\pi}{2r_h}=nk_h\\
k_h=\frac{2\pi}{\lambda_h}=\frac{\pi}{2r_h},\ \lambda_h=4r_h\nonumber
\end{eqnarray}
This leads to a discrete energy spectrum for the inside radial modes  which decrease the vacuum zero quantum fluctuation energy density inside.

\textbf{Zero energy density of vacuum outside.} Outside the horizon, the zero energy density of vacuum contributed from the zero point energy of one kind of spin states of the fermion-like modes can be calculated by virtue of the fermion gas model as follows. The vacuum zero-point energy outside in volume $V$ is
\begin{eqnarray}
(2\pi)^{-2}\hbar cV\int^{\pi k_p}_{0}k^3dk=\frac{\hbar cV}{4(2\pi)^2}(\pi k_p)^4
\end{eqnarray}
$\pi k_p$ is the fermion wave vector related to Planck wave vector. The zero energy density of vacuum outside is
\begin{eqnarray}
\rho_{p0}=\frac{\hbar c}{4(2\pi)^2}(\pi k_p)^4\\
k_p=\frac{1}{2r_p}=\frac{N_p}{2r_h},\ \lambda_p=4\pi r_p\\ N_p=\frac{r_h}{r_p}=\frac{10^5cm}{10^{-33}cm}\sim10^{38}
\end{eqnarray}

\textbf{Zero Energy density of vacuum inside.} Since the energy spectrum inside is discrete,
the integration should be replaced by summation and the fermion wave vector is
$k_N=N{\pi}/{2r_h}=N\pi k_p=N{\pi}/{2r_p}$,\ $N={r_h}/{r_p}=N_p$ . The results are
\begin{eqnarray}
\sum^{N-1}_{n=0}n^3=\frac{1}{4}{N(N-1)^2}\approx\frac{N^4}{4}(1-\frac{2}{N})= \frac{N_p^4}{4}(1-\frac{2}{N_p})\\
(2\pi)^{-2}\hbar cV(\frac{\pi}{2r_h})^4\sum^{N-1}_{n=0}n^3=\frac{\hbar cV}{4(2\pi)^2}(\frac{\pi}{2r_h})^4(N(N-1))^2 \approx\frac{cV}{4(2\pi)^2}(\pi k_p)^4(1-\frac{2}{N_p})
\end{eqnarray}

The zero energy density of vacuum inside is
\begin{eqnarray}
\rho_{h0}=\frac{\hbar c}{4(2\pi)^2}(\pi k_p)^4(1-\frac{2}{N_p})
\end{eqnarray}
The difference of zero energy densities of vacuum between outside and inside regions is\\
\begin{eqnarray}
\Delta\rho_{p0}=\rho_{p0}-\rho_{h0}=\frac{\hbar c}{4(2\pi)^2}(\pi k_P)^4[1-(1-\frac{2}{N_p})]=
  \frac{2\hbar c}{4(2\pi)^2}(\pi k_p)^4/N_p=2\rho_{p0}/N_p,\\
\Delta\rho_{p0}=\frac{2}{N_p}\rho_{p0},\ \rho_{p0}=\frac{\hbar c}{4(2\pi)^2}(\pi k_p)^4\\
\Delta\rho_{p0}/\rho_{p0}=\frac{2}{N_p}=\frac{2r_p}{r_h}= \frac{2m_p}{m_h}\sim\frac{10^{-33}}{10^5}=10^{-38}
\end{eqnarray}
$\Delta\rho_{p0}$ is the zero energy density loss of the inside vacuum from one kind of spin state of Planckon due to the cutoff boundary condition at the spherical horizon-the \textbf{Casimir effect.}\cite{7,8}

Since vacuum consists of densely piled Planckons and the vacuum energy density is the Planckon energy density $\rho_v=\rho_p$ £¬$\rho_{p0}=\rho_p/2$, and $e_{p0}=\rho_{p0}v_p$. Inside the black hole, each Planckon of one spin state loses zero energy
\begin{eqnarray}
\Delta e_{p0}=\Delta\rho_{p0}\nu_{p}=\frac{2e_{p0}}{N_p}=\frac{c\hbar}{2r_h}=e_h ,\\
e_{p0}=\frac{\hbar c}{4r_p},\ e_p=\frac{\hbar c}{2r_p},\ N_p=\frac{r_h}{r_p}\nonumber
\end{eqnarray}
%$e_{p0}=\frac{\hbar c}{4r_p}$, $e_p=\frac{\hbar c}{2r_p}$, $N_p=\frac{r_h}{r_p}$
which turns to be a radiation hole with exactly the minus mean energy \ $e_h$  of the radiation quantum of the spherical standing wave of the black hole.
 This in turn implies a temperature decrease $\Delta T$. According to the temperature Green's function for fermion\cite{7,8}, the mean energy $e_h$ and the temperature decrease $\Delta T$ has the relation
\begin{eqnarray}
\Delta e_{p0}/k_B\Delta T=\beta e_h=\pi,\ \pi k_B\Delta T=e_h=\frac{c\hbar}{2r_h}\\
\Delta T=\frac{\hbar c}{2\pi ck_Br_h}=\frac{\hbar \kappa}{2\pi ck_B}
\end{eqnarray}
where $\kappa={2GM}/{r_h^2}={c^2}/{r_h}$. This is exactly the Hawking-Unruh Formulae for the black hole, which is obtained from a different principle-the Casimir effect due to the geometric boundary condition for the inside vacuum excitation waves. Huge number of radiation quanta with mean energy \ $e_h$  and  moving in spherical horizon will produce the outward centrifugal energy flow and the corresponding centrifugal temperature $T_h=\frac{\hbar \kappa}{2\pi ck_B}(\kappa=\frac{c^2}{r_h})$ which balances the inward temperature decrease\ $\Delta T$  due to Casimir effect, namely $\Delta T=T_h$. Thus a thermodynamic equilibrium is established in the horizon.

   The above derivation of the black hole temperature tell us that the black hole temperature at the horizon stems from the mean energy difference of the excitation quanta in vacuum inside and outside the horizon, so that it is not the temperature itself, but the temperature difference of the vacuum quantum fluctuations inside and outside the horizon. From this sense, one would say that the Hawking-Unruh temperature of gravity  is  a vector quantity,  not a scalar quantity.

\subsection{Black hole mass: gravitation mass and physical mass}

The black hole absorbs particles from outside. The following calculations show that the absorbed particles are changed into spin 1/2 radiation quanta stored in the horizon and with mean energy \ $e_h=\hbar c/2r_h$ and the  mean zero energy $e_{h0}=\hbar c/4r_h$. The number of radiation quanta is exact the number of Planckons in the horizon which consists of only one Planckon layer with the thickness of the Planckon diameter\ $d=2r_p$. This means that each Planckon in the horizon has a resident radiation quantum and the host Planckon provides its position quantum number to the resident radiation quantum as a hidden quantum number. This position hidden quantum number resolves the Pauli exclusion problem for the super condensate in the horizon of the extremely large number of the fermion-type radiation quanta with extremely long wave length $\lambda_h=4\pi r_h$. The orbit of each spin 1/2 radiation quantum like a string forms a largest circle doubly rounded on the horizon sphere, all the radiation quantum orbits  in the horizon layer thus spin a densely string web, which is of the spin textile structure around the horizon sphere.

For the densely piled Planckon vacuum, The number\ $N_h$  of Planckons in the horizon layer is equal to the area of the sphere\ $4\pi r_h^2$   divided by the
area of the largest circle of the Planckon sphere \ $\pi r^2_p$:
\begin{eqnarray}
N_h=\frac{4\pi r_h^2}{\pi r_p^2}=4(\frac{r_h}{r_p})^2
\end{eqnarray}
The total mass\ $M_A$  of the zero energy mass of the radiation quanta in the horizon is $N_h$ times the radiation quantum zero energy mass $m_{h0}=e_{h0}/c^2=\frac{\hbar}{4cr_h}$,\\
\begin{eqnarray}
M_A=m_{h0}4(\frac{r_h}{r_p})^2/c^2=\frac{\hbar}{cr_h}(\frac{r_h}{r_p})^2
=\frac{\hbar r_h}{cr_p^2}=\frac{\hbar 2GM}{(G\hbar/c^3)c^3}=2M
\end{eqnarray}
which is twice of the black hole mass $M$,  half of which($M_A$) is used to compensate the  negative vacuum energy (negative gravity potential energy) mass.

To understand how the twice energy is formed, consider a particle from the infinite with the energy $\hbar\omega_0$  and reaching the black hole horizon with the energy\ $\hbar\omega(r_h)$ and mass $m(r_h)=\hbar\omega(r_h)/c^2$. The Newtonian Potential energy is $-\frac{\hbar\omega(r_h)}{c^2}\times\frac{GM}{r_h}$\ ($r$-dependence of mass $m(r)$ plays the role of general relativistic effect ).  According to energy conservation,  one has $\hbar\omega(r_h)-\frac{\hbar\omega(r_h)}{c^2}\times\frac{GM}{r_h}= \hbar\omega_0$ and $\hbar\omega(r_h)=\hbar\omega_0/(1-\frac{GM}{c^2r_h})= 2\hbar\omega_0$. So as the particle reaches the horizon, its energy and mass are doubled.

The same result is obtained for a particle at infinite with the rest energy\ $m_0c^2$  and at horizon with the energy $m(r_h)c^2:m(r_h)c^2-m(r_h)c^2\times\frac{GM}{c^2r_h}= m_0c^2$ and $m(r_h)c^2=m_0c^2/(1-\frac{GM}{c^2r_h})=2m_0c^2$. The energy and mass are also doubled as it reaches the horizon.

Since half of the energy $M_Ac^2$ is used to compensate the negative vacuum energy loss related to the negative gravity potential, the physical mass of the black hole is the residual mass $M$. However the gravitation mass $M_A=2M$  manifests itself as general relativistic effect in the Einstein potential $\phi(r)=-\frac{2GM}{r}$  and  in the related gravity strength $\kappa(r_h)=\frac{2GM}{r_h^2}$.

In what follows, let us answer the question where the lost energy of the vacuum inside black hole goes. As shown in sect. II-B, inside horizon each Planckon of one kind of spin states loses enegy $\Delta e_{p0}=e_h=e_p(\frac{r_p}{r_h})=2e_{p0}(\frac{r_p}{r_h})$ and mass $\Delta m_{p0}=\Delta e_{p0}/c^2$. In Planckon densely piled vacuum, the number of Planckons inside horizon is $N_v=(\frac{4\pi r_h^3}{3})/(\frac{4\pi r_p^3}{3})=(\frac{r_h}{r_p})^3$.  All the Planckons with one kind of spin states  lose energy and mass totally amounting to $E_T=N_v\Delta m_{p0}c^2$ and $M_T=N_v\Delta m_{p0}$:
\begin{eqnarray}
E_T&=&\Delta e_{p0}(\frac{r_h}{r_p})^3=2e_{p0}(\frac{r_h}{r_p})^2=\frac{1}{2}e_{p0}N_h,\nonumber\\
M_T&=&E_T/c^2=2m_{p0}(\frac{r_h}{r_p})^2=\frac{1}{2}m_{p0}N_h
\end{eqnarray}
Since the Planckon inside horizon has two spin states and  $N_h=4{r^2_h}/{r^2_p}$ is the number of Planckons in the spherical horizon layer, $2E_T$ and $2M_T$ contributed from two kinds of spin states of the inside Planckons are equal to the energy and mass of the Planckon layer with one kind of spin states. It will be showed in sect.III that the vacuum energy loss outside horizon which is dual to the inside one is also removed to the spherical horizon layer to form  another layer of Planckons with a different spin state.

\subsection{Microscopic origin of black hole entropy}

It is showed that black hole mass $M$ and energy $Mc^2$ come from radiation quanta in horizon, which are spin 1/2 fermions with the mean energy $e_h={\hbar c}/{2r_h}$ and the zero mean energy $e_{h0}={\hbar c}/{4r_h}$. The particles absorbed by the black hole have changed into radiation quanta and stored their information by virtue of microscopic states of radiation quanta. The number of radiation quanta equals the number of the Planckons in the horizon layer, namely $N_h=4{r_h^2}/{r_p^2}$.  Each radiation quantum has two states. The total number of microscopic states of the quantum many body system as an ideal gas with $N_h$  spin 1/2 fermions is $\Omega=2^{N_h}$.  According to the microscopic ensemble of quantum statistical physics, the entropy of the black hole is:
\begin{eqnarray}
S_h=k_Bln\Omega=k_Bln2\times N_h=k_Bln2\times\frac{4r_h^2}{r_p^2}=ln2\times\frac{k_BA_hc^3}{\pi\hbar G}=0.22\frac{k_BA_hc^3}{\hbar G}
\end{eqnarray}
where the black hole horizon area $A_h=4\pi r^2_h$ and $r^2_p={G\hbar}/{c^3}$. The above result is comparable with Hawking, Rovelli, and Shao as follows: Hawking\cite{4}: $S_h=0.25{k_BA_hc^3}/{\hbar G}$, Rovelli\cite{8}: $S_h={k_BA_hc^3}/{16\pi\hbar G}$, and Shao\cite{9}: $S_h=ln2\times{k_BA_hc^3}/{8\pi\hbar G}$.

\subsection{Thermodynamic equilibium and mechanical balance in horizon}

For the ideal gas of radiation quanta with mean energy $e_h$ in the horizon shell, there exists thermodynamic equilibrium and mechanical balance as follows. The inward Casimir-gravity temperature $\pi k_B\Delta T={\hbar \kappa}/{2c}={\hbar c}/2r_h(\kappa=2GM/r^2_h=c^2/r_h)$ and  the outward centrifugal temperature $\pi k_BT_{centrif}=e_h={\hbar c}/2r_h={\hbar \kappa}/{2c}(e_h=\hbar c/2r_h)$     are in thermodynanic equilibrium:
\begin{eqnarray}
T_{gr}=T_{centrif}(T_{gr}=\Delta T)
\end{eqnarray}
The gravitation acceleration $a_{gr}=\kappa={2GM}/{r_h^2}={c^2}/{r_h}$  and the centrifugal acceleration $a_{centrif}={c^2}/{r_h}$  are in mechanical balance:
\begin{eqnarray}
a_{gr}=a_{centrif},\ F_{gr}=ma_{gr}=F_{centrif}=ma_{centrif}
\end{eqnarray}
The above thermodynamic equilibrium and mechanical balance imply that only the radiation quanta with speed of light can exist and be tightly bound in the horizon, the non-radiation particles with speed less than c can not stay in the horizon stably. For this kind of radiation quanta, the outward centrifugal energy flow is balanced by the inward gravity energy flow, so that the radiation quanta in the horizon are in thermodynamic equilibrium. The thermodynamic equilibrium and mechanical balance make the vacuum space-time of the horizon area singular, and the horizon surface becomes a cutoff boundary for the interior radial wave modes.

\subsection{Gravitation effect of inside radiation quanta and attractor behaviour of horizon}

We have shown that inside the black hole, each Planckon loses energy and creates a radiation quantum hole with mean energy $e_h$ which
corresponds to an  constant temperature decrease $\pi k_B\Delta T=e_h={\hbar c}/{2r_h}={\hbar \kappa}/{2c}$  and a gravitation acceleration
$\kappa={c^2}/{r_h}$. Since the radiation quantum hole has negative energy, from $\kappa= -\frac{d\phi(R)}{dR}$ one gets the negative Einstein potential $\phi(R)=-{c^2}R/{r_h}$, which is linear in $R$ and quite different from the conventional inside potential $\phi(R)=-{c^2r_h}/{R}(R<r_h)$, which is singular at the origin $R=0$. From $\phi(R)$, we find that inside the black hole, no singularity exists at $R=0$ and its radial acceleration points to the horizon surface.  It is well known that outside the horizon, the
radial gravitation acceleration also points to the horizon surface. Thus the two-side gravity forces make the horizon surface become a
special surface with attractor behavior. In Appendix A,  it is shown that there is a dual relation between the inside and outside gravity potentials.
From the dual transformation $R={r_h^2}/{r}$, one can obtain the outside Einstein potential $\phi(r)=-\frac{2GM}{r}$  from the inside
one $\phi(R)$. The dual relation indicates that based on the physics of Casimir effect, the inside gravity potential of the Schwarzschild black hole  has a completely different physical behavior in contrast to the conventional conjecture.

\section{Microscopic structure of Einstein potential outside}

In the last subsection, we have obtained the Einstein potential outside \ $\phi(r)$  from the inside one which is obtained microscopically by using quantum statistical physics of ideal radiation quantum gas. In this section, we shall derive this potential by Laplace equation for local equilibrium temperature\cite{11}.

From the above discussion we know that the gravitation temperature from the negative Einstein potential corresponds to a radiation quantum hole with
minus mean energy $e(r)$  which is related to gravity potential energy $\Phi(r):-\Phi(r)=e(r)$  and satisfies the period condition of temperature Green's function for fermion:
\begin{eqnarray}
e(r)/k_B\Delta T(r)=\beta(r)e(r)=\pi
\end{eqnarray}
or
\begin{eqnarray}
\Phi(r)/\pi=-k_B\Delta T(r)
\end{eqnarray}
The Laplace equation for local equilibium temperature is \cite{11}:
\begin{eqnarray}
{\nabla^2}\Phi(r)= - \pi k_B{\nabla^2}\Delta T(r)=0
\end{eqnarray}
The solution is
\begin{eqnarray}
\Phi(r)=\frac{C}{r}
\end{eqnarray}
From the boundary condition at $r_h:\Phi(r_h)={C}/{r_h}$   and
\begin{eqnarray}
\Phi(r_h)/\pi=-k_B\Delta T(r_h)=-\frac{\hbar \kappa}{2\pi c}=-\frac{\hbar c}{2\pi r_h}\\
\kappa=\frac{2GM}{r^2_h}=\frac{c^4}{2GM}=\frac{c^2}{r_h}
\end{eqnarray}
 one obtains\  $\Phi(r_h)=-e(r_h)=-{\hbar c}/{2r_h}$ ,\ $C=-\hbar c/2$ , and
\begin{eqnarray}
\Phi(r)=-\frac{\hbar c}{2r},\ e(r)=\frac{\hbar c}{2r},\ k_B\Delta T(r)=k_B\Delta T(r_h)\frac{r_h}{r}
\end{eqnarray}
However $\Phi(r)$ is not Einstein potential $\phi(r)=-\frac{2GM}{r}$, instead it is the gravity potential energy of a radiation quantum with mass $m_h=\frac{\hbar}{2cr_h}$ and energy $e_h=\frac{\hbar c}{2r_h}$: $\Phi(r)=-\frac{2GMm_h}{r}=-\frac{\hbar c}{2r}=-e(r)$ .  So the Einstein potential outside should be:
\begin{eqnarray}
\phi(r)=\frac{\Phi(r)}{m_h}=-\frac{c^2r_h}{r}=-\frac{2GM}{r}
\end{eqnarray}
 We have shown that inside the horizon, the Casimir effect leads to vacuum energy loss and creation of the spin 1/2 radiation quantum hole with the negative value of the mean energy $e_h=\frac{\hbar c}{2r_h}$  which is related to the temperature and gravity strength by the relation:
\begin{eqnarray}
e_h/\pi=k_BT(r_h)=\frac{\hbar\kappa(r_h)}{2\pi c},\ (\kappa(r_h)=\frac{2GM}{r^2_h}=\frac{c^2}{r_h})
\end{eqnarray}
Similarly, outside horizon, the negative gravitation potential $\phi(r)$ and the gravitation strength $\kappa(r_\kappa)=\frac{2GM}{r^2}=\frac{c^2}{r_\kappa}$ also lead to vacuum energy loss and creation of the spin 1/2 radiation quantum hole with the negative value of the mean energy
\begin{eqnarray}
\label{3-7}
e(r_\kappa)=\frac{\hbar c}{2r_\kappa},\ (r_\kappa=r\frac{r}{r_h})
\end{eqnarray}
which is related to the temperature and gravity strength by the relation
\begin{eqnarray}
\label{3-8}
e(r_\kappa)/\pi=k_BT(r_\kappa)=\frac{\hbar \kappa(r_\kappa)}{2\pi c},\
\kappa(r_\kappa)=\frac{c^2}{r_\kappa},\ T(r_\kappa)=\frac{\hbar c}{2\pi k_Br_\kappa}
\end{eqnarray}
This is the generalized Hawking-Unruh formulae from inside to outside regions.

Outside the horizon, let us consider the radiation quantum as a standing wave moving on  $r$-sphere. This radiation quantum will have the mean energy\  $e(r)={\hbar c}/{2r}$, with the corresponding gravitation acceleration \ $\kappa(r_\kappa)={2GM}/{r^2}={c^2}/{r_\kappa}$ and gravitation temperature\ $T(r_\kappa)=\frac{\hbar c}{2\pi k_Br_\kappa}$. Its centrifugal acceleration is\ $a_{centrif}=\frac{c^2}{r}$  and centrifugal temperature is $T(r)=\frac{\hbar c}{2\pi k_Br}$ . Since $\frac{c^2}{r_\kappa}<\frac{c^2}{r}$  and \ $T(r)>T(r_\kappa)$ , the radiation quantum \ $e(r)$  can not stay in\ $r$-sphere stably, it will leave $r$-sphere.  From the generalized relations Eq.(\ref{3-7} - \ref{3-8}) and the above consideration, we find that there exist thermodynamic non-equilibrium and mechanical non-balance on the $r$-sphere, namely the inward gravitation temperature \ $T(r_\kappa)$ is smaller than the outward centrifugal temperature \ $T(r)$, and the centrifugal acceleration \ $a_{centrif}(r)=c^2/r$  is larger than the gravitation acceleration\ $a_{gr}(r_\kappa)=\kappa(r_\kappa)=\frac{c^2}{r_\kappa}$ . These thermodynamic non-equilibrium and mechanical non-balance lead to two kinds of radiation quantum energy flows with opposite directions: the outward centrifugal driving energy flow and the inward gravitation driving energy flow, their compensation establishes local dynamical energy equilibrium. These processes and flow lines should be studied in detail by using the geodesic equations for zero geodesic trajectories of the radiation quanta.

Now, let us answer the question where the lost energy of vacuum outside the black hole goes.  Outside the horizon, the Einsten gravity potential and its gravitation acceleration are:
\begin{eqnarray}
\phi(r)=-\frac{2GM}{r}=-\frac{c^2r_h}{r},\ \kappa(r_\kappa)=\frac{2GM}{r^2}=\frac{c^2r_h}{r^2}=\frac{c^2}{r_\kappa}\nonumber\\
r_\kappa(r)=r(\frac{r}{r_h})>r
\end{eqnarray}
We have shown in Eq.(\ref{3-7} - \ref{3-8}) that outside the horizon,
the energy loss of vacuum creates the spin 1/2 radiation quantum
hole with the negative value of the mean energy\ $e(r_\kappa)={\hbar c}/{2r_\kappa}$ ( its
zero energy is \ $e_0(r_\kappa)={\hbar c}/{4r_\kappa}$ ) and induces the
statistical temperature\ $T(r_\kappa)$. Since the radiation quanta
in the $r$-sphere shell and in the horizon sphere shell are radially
projectively connected, the number of the radiation quanta in the $r$-sphere
shell is the same as that in the horizon sphere shell, namely the number of
quanta is also\ $N_h=4(\frac{r_H}{r_p})$. The differential total
zero energy of the radiation quanta in the $r$-sphere shell with the thickness
\ $d=2r_p$ is
\begin{eqnarray}
\label{3-10a}
dE(r)=e_0(r_\kappa)N_h\frac{dr}{2r_p}=\frac{\hbar c}{8r_\kappa}\times 4(\frac{r_h}{r_p})^2=\frac{1}{2}\hbar c(\frac{r_h}{r_p})^3\frac{dr}{r^2}
\end{eqnarray}
To obtain the total zero energy of all the radiation quanta with one kind of spin states outside the horizon,
Eq.(\ref{3-10a}) should be integrated from $r_h$ to $\infty$,
\begin{eqnarray}
E_T=\int^\infty_{r_h}dE(r)=\frac{1}{2}\hbar c(\frac{r_h}{r_p})^3\int_{r_h}^\infty\frac{dr}{r^2}=\frac{\hbar c}{8r_p}4(\frac{r_h}{r_p})^2=\frac{1}{2}e_{p0}N_h
\end{eqnarray}
Since each radiation quantum has two spin states, $2E_T$ is exactly the zero energy of the Planckons with one kind of spin states in the sphere layer at $r_h$. Thus outside the horizon, the vacuum energy loss  manifesting itself as both negative gravitation potential and radiation quantum hole energy $-e_0(r_\kappa)$,  has been removed to the horizon Planckon layer with one kind of spin states, just like the inside vacuum energy loss manifesting itself as radiation quantum hole energy  $-e_h$ has also been removed to the same layer with the opposite spin state. In appendix A , by a dual transformation, we do the same calculation and produce the same result.

\section{Gravitation potential of black hole in accelerating universe}

For the accelerating universe, the temperature depends on time in the inverse way of the universe radius: $T(r,t)=T(r)e^{-\Lambda t}$, where $\Lambda$ is the measure of expansion rate.

 For the time-dependent temperature,  its evolution equation reads\cite{11}:
\begin{eqnarray}
\frac{\partial\Delta T}{\partial t}-\frac{k}{\rho c_p}\nabla^2\Delta T=0
\end{eqnarray}
where $c_p$ is the specific heat with fixed volume, which is negative for black holes, $k$ is heat conductance.  Since $\Delta T \sim \frac{1}{R(t)}\sim e^{-\Lambda t}$  and $k_B\Delta T\sim -\Phi$,  as $r>r_h$ we have the equation for the gravitation potential energy as follows :
\begin{eqnarray}
\nabla^2\Phi(r)+\frac{\rho c_p \Lambda}{k}\Phi=0,\nonumber\\
\nabla^2\Phi(r)-\frac{1}{\lambda^2}\Phi=0,\\
\lambda=\lambda_{cosmon}=(-\frac{\rho c_p \Lambda}{k})^{-\frac{1}{2}}\nonumber
\end{eqnarray}
 where the mass term is induced in the stationary equation and the corresponding new particle is called cosmic expansion cosmon, with its  Compton wave length $\lambda_{cosmon}\approx 10^{28}cm$, and its energy and mass :
\begin{eqnarray}
e_{cosmon}=\frac{2\pi\hbar c}{\lambda}\approx10^{-45}erg,\ m_{cosmon}=\frac{2\pi \hbar}{\lambda c}=10^{-66}g,\ m_{cosmon}/m_p\approx 10^{-61}
\end{eqnarray}
As the universe expanding, the space-time symmetry of the time dependent temperature equation is broken, the gravity excitation quantum acquires mass $m_{cosmon}$.

In viewing the boundary condition $\Phi(r_h)=-\frac{2GMm}{r_h}e^{-r_h/\lambda}$, the solution of the stationary potential energy equation is $\Phi(r)=-\frac{2GMm}{r}e^{-r}/\lambda$,\ and the Einstein potential  is
\begin{eqnarray}
\phi(r)=\Phi(r)/m=-\frac{2GM}{r}e^{-r/\lambda}=-\frac{2GM}{r}+\frac{2GM}{r}(1-e^{-r/\lambda}) \approx -\frac{2GM}{r}+\frac{2GM}{\lambda}
\end{eqnarray}
(For the solar mass, $2GM_\odot/\lambda\sim10^{-2}(cm/s)^2\sim10^{-22}c^2$)

If the cosmic expansion cosmon has the Planckon cross section $\sigma=(\frac{\hbar}{m_p c})^2$ ( the cosmon is the collective excitation of huge number of Planckons ) and  its volume is $V_{cosmon}=\lambda_{cosmon}\sigma$,  the mass density of the cosmic expansion cosmon is thus
\begin{eqnarray}
\label{4-4}
\rho_{cosmon}=\frac{m_{cosmon}}{\frac{\hbar}{m_{cosmon}c}(\frac{\hbar}{m_p c})^2}= \frac{m^2_{cosmon}}{m_p(\frac{\hbar}{m_p c})^3}=\frac{m_p}{(\frac{\hbar}{m_pc})^3}\times 10^{-122}\sim\rho_v\times 10^{-122}=\rho_\Lambda
\end{eqnarray}
where $m_{cosmon}/m_p\approx10^{-61}$,\ $\rho_p=\rho_v\sim {m_p}/{(\frac{\hbar}{m_pc})^3}$ is the mass density of Planckon or vacuum, and $c^2 \rho_\Lambda$ is just the dark energy density. Eq.(\ref{4-4}) indicates that if the universe is fully filled by the cosmons, the energy density of the fully filled cosmic expansion cosmons is in the same order of magnitude as the dark energy density. The above result is consistent with the universe expansion model with the Planck era as the initial condition of the universe\cite{12}.

\section{Microscopic quantum statistical structure of gravity in general}

 For general gravitation fields, the Einstein potential $\phi(\vec{r})$  is  a  functional of metrices $g_{\mu\nu}(r)$. If the functional is known,
\begin{eqnarray}
\phi(\vec{r})=\phi[g_{\mu\nu}(\vec{r})]
\end{eqnarray}
the gravitation acceleration can be calculated from the Einstein potential:
\begin{eqnarray}
\vec{\kappa}(\vec{r})=-\nabla\phi(\vec{r})
\end{eqnarray}
On the other hand, for a system with spherical symmetry or for a circular orbit, one can calculate the acceleration from zero geodesic trajectory of the quantum in semi-classical approximation,
\begin{eqnarray}
\kappa(r)=\sqrt{g_{rr}g_{00}}\frac{d^2r}{d\tau^2}=c^2\sqrt{\frac{g^{rr}}{g_{00}}}\nabla_rg_{00}
\end{eqnarray}
From mechanical balance $\kappa=\frac{c^2}{r_\kappa}$, we obtain the gravitation curvature radius  $r_\kappa=\frac{c^2}{\kappa}$ and the mean energy $e(r_\kappa)=\frac{\hbar c}{2r_\kappa}=\frac{\hbar \kappa}{2c}$ of the corresponding radiation quantum hole which is a standing wave on the  $r_\kappa$-sphere with wave length  $\lambda_\kappa=4\pi r_\kappa$   and wave vector $k_\kappa=\frac{1}{2r_\kappa}$. From the temperature Green's function\cite{7,8}, $\beta(r_\kappa)e(r_\kappa)=e(r_\kappa)/k_BT(r_\kappa)=\pi$, one obtains the temperature of the gravitation system,
  \begin{eqnarray}
k_BT(\vec{r}_\kappa)=e(r_\kappa)/\pi=\frac{\hbar c}{2\pi r_\kappa}=\frac{\hbar \kappa}{2\pi c}=\frac{\hbar}{2\pi c}|\vec{\nabla}\phi[g_{\mu\nu}(\vec{r})]|=\frac{\hbar c}{2\pi}\sqrt{\frac{g^{rr}}{g_{00}}}\nabla_rg_{00}.
\end{eqnarray}
The last formulae in the above equation is for the Schwarzschild balck hole.

Till now, the key information of the microscopic quantum statistical structure of the gravitation system is known: the microscopic constituents of the system are radiation quantum holes with spin 1/2 and negative mean quantum energy $-e(r_\kappa)=-\frac{\hbar c}{2r_\kappa}$, the local statistical temperature
( decrease ) is  $T(r_\kappa)$. In general $T(r_\kappa)$ is local and of space-time dependence, which leads to the non-thermodynamic equilibrium and non-mechanical balance, as well as the related energy flows with opposite directions  as discussed in sect.III.  Since the Planckon vacuum structure is known, the number( or number density ) of constituents depends on the gravitation field distribution and the geometric configuration of the system as shown in the black hole case and can be calculated in a similar way as the black hole case.

\section{Summary and discussion}

In this article, we have proposed the Planckon densely piled model of vacuum. Based on this vacuum model, the microscopic quantum structure of the Schwarzschild black hole and the quantum statistical origin of its gravity have been studied.  It is shown that the thermodynamic temperature equilibrium and the mechanical acceleration balance make the space-time of the black hole horizon become a singular surface,  only radiation particles can exist stably and be confined tightly in the horizon sphere, moving in the horizon with the speed of light. For this singular surface, the Casimir effect works inside the horizon. It makes the inside vacuum have less zero quantum fluctuation energy than the outside vacuum, the temperature difference( outside $T_{out}$ high, inside $T_{in}$ low) and the gravity as its thermal pressure are created. Inside the black hole, each Planckon of the vacuum loses zero quantum fluctuation energy an amount of $e_{h}={\hbar c}/{2r_h}$, the lost vacuum zero quantum fluctuation energy manifests itself as the gravitation strength $\kappa=\frac{2GM}{r_h^2}=\frac{c^2}{r_h}$  and the corresponding negative Einstein potential $\phi(R)=-\kappa R$, which is is linear in radial coordinate and without singularity at the origin of the black hole, in sharp contrast to the conventional view of point. Then a dual relation between inside and outside gravity potentials of the black hole is found.  By the dual transformation $R\leftrightarrow\frac{r_h^2}{r}$ , the interior potential $\phi(R)=-\kappa R(R<r_h)$  is transformed into the exterior potential $\phi(r)=-\frac{2GM}{r}(r>r_h)$. The lost vacuum energy in the negative gravitation potential regions has been removed to the black hole horizon surface to form a spherical Planckon shell with the thickness $d=2r_p$ .  All the particles absorbed by the black hole have fallen down to the horizon and converted into radiation quanta consisting of standing waves on the  $r_h$-sphere with the spin 1/2 and the mean energy $e_h={\hbar c}/{2r_h}$ ,  the thermodynamic equilibrium and the mechanical balance keep them stable and be tightly bound in the horizon. All the orbits of radiation quantum holes in the horizon spin a densely string web around the horizon. The total mass of the radiation quanta in the horizon becomes the gravitation mass $2M$, half of which is used to compensate the negative gravitation potential energy mass, the physical mass of the black is thus $M$. The entropy of the black hole has been calculated from the microscopic state number of the quantum many-body system  consisting of ideal radiation fermion gas and the result is well consistent with  Hawking.   Outside the horizon, there exist a thermodynamic non-equilibrium and a mechanical non-balance which lead to an outward centrifugal driving energy flow and an inward gravitation driving energy flow, their compensation establishes the local dynamical energy equilibrium. The four thermodynamic laws of the black hole can be understood in terms of quantum statistical physics.

In this paper, the derivation of the black hole temperature  tells us that the black hole temperature at the horizon is resulted from the mean energy difference and the corresponding temperature difference of the excitation quanta in vacuum inside and outside the horizon,  so that  one would say that the Hawking-Unruh temperature of gravity  is  not a scalar quantity, rather it is a vector quantity, since it is of temperature difference. This in turn implies that the gravity phenomena is related to global non-equilibrium( local equilibrium ) processes  in the sense of thermodynamics and statistical physics.

The accelerating expansion of the universe yields the expansion cosmon and its energy density agrees  with dark energy density  in the order of magnitude.

The results for the black hole have been generalized to general gravitation cases, and the procedures how to find the microscopic quantum statistical structure of a gravitation system are discussed.

The study and results of this article are quite preliminary. The Planckon piled vacuum model is constructed quantum-mechanically at the mean field and semi-classical level, since the basic brick of vacuum-the Planckon is introduced as a semi-classical quantum object ( a standing quantum wave in the Planckon $r_p$-sphere with the wave length $\lambda =4\pi r_{p}$ ) and quantum fluctuations around mean values are neglected. The quantization of the black hole excitations is also at the semi-classical level, namely the quantization of vacuum excitations is based on quantized orbital standing waves, just like that N.Bohr had quantized electron motion in the hydrogen atom as an quantized orbital standing wave. However, in our case, the spherical black hole plays the role of the hydrogen atom. Furthermore, as a quantum statistical system of gravity, the radiation excitation quanta of vacuum are treated as an ideal fermion gas at the temperature-dependent mean field, and the interactions among them are not included. In a mean field treatment of a quantum statistical many-body system, quantum and statistical fluctuation around the mean value is not considered. There are two kinds of fluctuations around the mean values: 1) the quantum mechanical fluctuation around the quantum mechanical mean value and  2) the statistical ensemble fluctuation around the ensemble average.  As the statistical mean energy of the radiation quanta of Fermion type, it is naturally related to a temperature by the period condition of the temperature Green's function: $e_h/\pi=k_BT_h$. As a result, the canonical ensemble of an ideal Fermion gas leads to the Dirac probability distribution $ f (\frac{e-\mu}{k_BT_h})=\frac{1}{e^{(e-\mu)/k_BT_h}+1}$  with $T_h=\frac{\hbar c}{2\pi k_Br_h}$ for the radiation fermion quanta. To consider the quantum zero energy fluctuation of the Planckon, the Planckon sphere ( the semi-classical description of the Planckon ) should be replaced by a Gaussian wave function of the zero energy quantum state of the Planckon (the full quantum mechanical description of the Planckon ) in the harmonic oscillator approximation for the small amplitude oscillation, and an effective temperature should be introduced with $k_BT_P\sim e_P/\pi$  for the energy distribution of the Planckon. In the fully quantum mechanical description, the quantum vacuum is fully filled by the Gaussian wave functions of Planckons and the universe vacuum space becomes quantum mechanically stochastic space.

In the semi-classical description of the vacuum, the radiation quanta correspond to extremely long classical trajectories and low energies, the number of radiation quanta is extremely large, both fluctuations over their mean values are negligibly small. That is why we obtain the exact quantities  for the black hole.

In our model, vacuum is a kind of Planckon densely piled crystal,  its radiation excitations are like phonons and its particle excitations are like dislocations or defects in solid\cite{6}. Thus the investigation of the relation between gravity and solid states is very interesting and significant\cite{13}. In this respect,  it has been shown that vacuum is a kind of super fluid medium\cite{14}.
It should be noted that our model may be related to the super-string theory\cite{15} and the loop quantum gravity theory\cite{9} in some respects. The detailed relation between our microscopic model and the macroscopic theory of general relativity is quite worthy to explore and the generalization of this study to different kinds of black holes  is  quite desirable.

\appendix

\section{Dual relation between inside and outside gravity potentials of Schwarzschild black hole}
Dual relation of coordinates:
\begin{eqnarray}
R\leftrightarrow r,\ R=\frac{r^2_h}{r}
\end{eqnarray}

Dual relation of inside and outside regions :
\begin{eqnarray}
r\subset[r_h,\ \infty) \leftrightarrow R\subset[0,\ r_h]
\end{eqnarray}

Dual relation between inside and outside Einstein potentials :
\begin{eqnarray}
\phi(r)=-\frac{2GM}{r}\leftrightarrow-\frac{c^2R}{r_h}=\phi(R)
\end{eqnarray}
It is remarkable that inside the black hole the Einstein gravity
potential is linear in  radial coordinate and no singularity exists
at the origin of the black hole in sharp contrast to the conventional view of
point.

 Dual relation of gravity strengths: outside
$\kappa(r)$ pointing to the horizon
\begin{eqnarray}
\kappa=-\frac{d\phi(r)}{dr}=-\frac{2GM}{r^2}=-\frac{c^2}{r_\kappa},\ r_\kappa=r(\frac{r}{r_h})
\end{eqnarray}
inside $\kappa(R)$  also pointing to the horizon
\begin{eqnarray}
\kappa(R)=-\frac{d\phi(R)}{dR}=\frac{c^2}{r_h}
\end{eqnarray}
The horizon surface thus behaves as an attractor surface!

The problem of the transport of the exterior vacuum energy loss ( related
 to the existence of negative potential energy ) to the surface can be converted into the interior problem. By virtue of the dual transformation, the exterior gravity potential and its strength can be transformed into interior ones. From the formulae $\pi k_B\Delta T=\frac{\hbar \kappa(R)}{2c}=\frac{\hbar c}{2r_h}=e_h\ (\kappa(R)=\frac{c^2}{r_h})$, we know that every Planckon inside vacuum loses energy $e_h$. The total number of Planckons in this region is  $N_T=(\frac{r_h}{r_p})^3$, so the total energy removed from this region to the surface is : $E_T=e_h(\frac{r_h}{r_p})^3=\frac{1}{2}\frac{\hbar c}{4r_p}4(\frac{r_h}{r_p})^2=\frac{1}{2}e_{p0}N_h$ , where  $N_h=4(\frac{r_h}{r_p})^2$ is the number of Planckons in the shell of black hole horizon surface and $e_{p0}=\frac{\hbar c}{4r_p}$ is Planckon zero energy. The result is the same as that obtained from the direct calculation in section III.

Acknowledgement: This work was supported in part by the National Natural Science Foundation of China under the grant No. 10974137 and 10775100, the Doctoral Education Fund of the Education Ministry of China, and by the Research Fund of the Nuclear Theory Center of HIRFL of China.

\end{document}